\documentclass[a4paper,12pt,fleqn]{article}
\usepackage{geometry}
\usepackage{amsmath}
\begin{document} 
\begin{center}
\textbf{\large
SPLITTING OF 3D QUATERNION DIMENSIONS INTO 2D-SELLS AND A ``WORLD SCREEN TECHNOLOGY''}
\\[1em]
Alexander P. Yefremov \\[0.5em] Institute of Gravitation and Cosmology of Peoples' Friendship University of Russia
\\ Miklukho-Maklaya str., 6, 117198, Moscow, Russia
\end{center}
\textbf{Abstract}
\\[0.3em]
{\small A set of basic vectors locally describing metric properties of an arbitrary 2-dimensional (2D) surface is used for construction of fundamental algebraic objects having nilpotent and idempotent properties. It is shown that all possible linear combinations of the objects when multiplied behave as a set of hypercomples (in particular, quaternion) units; thus interior structure of the 3D space dimensions pointed by the vector units is exposed. Geometric representations of elementary surfaces (2D-sells) structuring the dimensions are studied in detail. Established mathematical link between a vector quaternion triad treated as a frame in 3D space and elementary 2D-sells prompts to raise an idea of ``world screen'' having 1/2 of a space dimension but adequately reflecting kinematical properties of an ensemble of 3D frames.}
\\[0.8em]

\vskip 1em
\section*{\small 1. INTRODUCTION}
\vskip-0.7em
Raise of interest to the mathematics of hypercomplex (HP) numbers has been definitely manifested in the last decade, e.g.$^{1-4}$. It was provoked not only by a ``dissipation of energy'' in aging physical theories of the gone century but rather by growing understanding of deep, yet not completely revealed, content of the mathematics offering a fruitful field for theoretical thought. One physical domain obviously linked with HP numbers is quantum theory with associative but often non-commuting operators that resemble objects from the bi-quaternion (BQ) set$^{5,6}$ . On the other hand the set is well known to comfortably represent main relations of theories of relativity$^{7,8}$. This pure mathematical relationship of still incompatible quantum physics and general relativity (gravitation) may one day make a path to reconcile the theories. Whether (or not) this occur, the detailed study of HP algebras already now demonstrate its benefit allowing to find many physical correlations in this remarkably rich mathematical medium$^{9}$. Here the study of basic elements of HP-numbers set is continued concerning geometric introduction of quaternion\footnote{Biquaternion and quaternion algebras have the same set of units, hence, associated spinors; so for simplicity further on the objects will be referred to only as Q-units and Q-spinors.} (Q-) spinors.

\noindent BQ numbers, a subset of HP numbers, may be regarded unique since it comprises diverse representatives of associative algebras, of ``good'' ones: real, complex and quaternion numbers, as well as of ``not good'' ones, admitting zero divisors: biquaternions, double (split complex) and dual numbers$^{10}$. In particular one will see below that the HP-unit of dual numbers despite the fact that it has zero norm is the ``most fundamental'' among all other units, the real one included. But a genuine elementary object lying in the very basement of all units forming the named above algebras is a set of spinors, or elements of ideals$^{11}$, that heuristically emerge as eigenfunctions of imaginary Q-units represented by square matrices$^{12}$. This way of introduction of the spinors suffers at least of two disadvantages. First, to hit the goal one has to use a definite representation of Q-units thus depriving the result of generality. Second, the character of the Q-spinors as (hidden) basic elements of the Q-algebra having transparent geometric meaning$^{13}$ comes to light in solution of equations for eigenfunctions of an operator, what in fact is an artificial mathematical act. It seems logically sustainable to start, vice versa, from elementary geometric notions and objects arriving in result to composed objects describing other geometry. This line is pursued in this study; its result is an original model of nature of 3D-world dimensions.

\noindent In Section 2 a geometric basement of the theory, a 2D surface locally described by a dyad (two unitary and orthogonal vectors), is introduced, and all primitive direct products of the basic vectors, nilpotent and idempotent objects, are investigated. Section 3 contains a detailed study of algebraic properties of all possible linear combinations of the nilpotent and idempotent objects. It is shown in Section 4 that the four obtained new objects form a set of vector HP units, in particular the set of Q-units, what prompts to regard the initial surface, 2D-sell, as a fundamental structural element of a 3D space dimension. In Section 5 the dyad vectors structuring the other dimensions of the space are found as functions of the initial 2D-sell elements, and expressions for respective metric tensors are deduced. An idea of ``world screen'' mathematically equivalent to a number of Q-frames (particles) in 3D space is suggested in Section 6, and samples of simple 2D-sells are considered giving birth to different Q-frames. Short discussion in Section 7 concludes the study.
\section*{\small 2. TENSOR PRODUCTS OF ORTHONORMAL VECTORS ON A SURFACE}
\vskip-0.7em
\noindent Consider a sufficiently smooth 2D space (surface) endowed with a coordinate system $x^{A} =\{ x^{1} ,x^{2} \} $ and having a symmetric metric $g_{AB} $, with its reciprocal existing $g^{BC} $: $g_{AB} g^{BC} =\delta _{A}^{C} $ ($A,\, B,\, C\, ...=1,2$, $\delta _{A}^{C} $ is the symbol of Kronecker, summation over repeated indices is implied). A square line element on the surface is 
\begin{equation} \label{GrindEQ__1_} 
ds^{2} = g_{AB} dx^{A} dx^{B}.   
\end{equation} 
In general the metric may be not Euclidian, so the difference between covariant and contravariant components of geometrical objects on the surface is important. It is always possible to select a couple of orthonormal vectors $a^{A} ,\, \, b^{B} $ in a fixed point of the surface:
\begin{equation} \label{GrindEQ__2_} 
g_{AB} a^{A} a^{B} =g_{AB} b^{A} b^{B} =1,  
\end{equation} 
\begin{equation} \label{GrindEQ__3_} 
g_{AB} a^{A} b^{B} =a^{A} b_{A} =0.  
\end{equation} 
Now axiomatically introduce tensor (direct) products of the vectors with components of mixed covariance, thus obtaining $2\times 2$-matrices
\begin{equation} \label{GrindEQ__4_} 
E^{A}_{\;\; B} \equiv a^{A} g_{BC} b^{C} =a^{A} b_{B} ,  
\end{equation} 
\begin{equation} \label{GrindEQ__5_} 
F^{A} _{\;\;B} \equiv b^{A} g_{BC} a^{C} =b^{A} a_{B} .  
\end{equation} 
The objects (\ref{GrindEQ__4_}), (\ref{GrindEQ__5_}) are non-symmetric matrices, due to Eq.(\ref{GrindEQ__3_}) they are traceless 
\[Tr(E^{A} _{\;\;B} )=E^{A} _{\;\;A} =Tr(F^{A} _{\;\;B} )=F^{A} _{\;\;A} =0,\] 
and have vanishing determinants 
\[\det (E^{A} _{\;\;B} )=\frac{1}{2} (\delta _{A}^{B} \delta _{C}^{D} -\delta _{A}^{D} \delta _{C}^{B} )(a^{A} b_{B} )(a^{C} b_{D} )=0, \quad \det (F^{A} _{\;\;B} )=0.\] 
Examine multiplication properties of the objects (\ref{GrindEQ__4_}) and (\ref{GrindEQ__5_}). Altogether one can form only four products. The first two products are squares of the matrices; the squares vanish due to Eq.(\ref{GrindEQ__3_})
\begin{equation} \label{GrindEQ__6_} 
E^{2} \equiv E^{A} _{\;\;B} E^{B} _{\;\;C} \equiv a^{A} b_{B} a^{B} b_{C} =0,   
\end{equation} 
\begin{equation} \label{GrindEQ__7_} 
F^{2} \equiv F^{A} _{\;\;B} F^{B} _{\;\;C} \equiv b^{A} a_{B} b^{B} a_{C} =0 ,  
\end{equation} 
hence \textit{E} and \textit{F} belong to the set of nilpotent matrices (or just nilpotents). The second two products of the nilpotents (\ref{GrindEQ__4_}), (\ref{GrindEQ__5_}) are result of their mixed left and right multiplication
\begin{equation} \label{GrindEQ__8_} 
G_{C}^{A} \equiv E^{A} _{\;\;B} F^{B} _{\;\;C} \equiv a^{A} b_{B} b^{B} a_{C} =a^{A} a_{C} ,  
\end{equation} 
\begin{equation} \label{GrindEQ__9_} 
H_{C}^{A} \equiv F^{A} _{\;\;B} E^{B} _{\;\;C} \equiv b^{A} a_{B} a^{B} b_{C} =b^{A} b_{C} .  
\end{equation} 
The objects (\ref{GrindEQ__8_}), (\ref{GrindEQ__9_}) are symmetric matrices with also vanishing determinants, but their traces equal unity due to Eq.(\ref{GrindEQ__2_})
\[\det (G_{B}^{A} )=\det (H_{B}^{A} )=0, \quad Tr(G_{B}^{A} )=G_{A}^{A} =Tr(H_{B}^{A} )=H_{A}^{A} =1.\] 
Examine multiplication properties of the objects (\ref{GrindEQ__8_}) and (\ref{GrindEQ__9_}). Again one can compose only four products. The first two products are the squares
\begin{equation} \label{GrindEQ__10_} 
G^{2} \equiv G_{B}^{A} G_{C}^{B} \equiv a^{A} a_{B} a^{B} a_{C} =G_{C}^{A} ,   
\end{equation} 
\begin{equation} \label{GrindEQ__11_} 
H^{2} \equiv H_{B}^{A} H_{C}^{B} \equiv b^{A} b_{B} b^{B} b_{C} =H_{C}^{A} .   
\end{equation} 
Eqs.(\ref{GrindEQ__10_}), (\ref{GrindEQ__11_}) state that the square of each matrix is the same matrix; therefore any natural power \textit{P} of the matrix returns itself
\[G^{P} =G; H^{P} =H;\] 
such algebraic objects are called idempotent matrices (or simply idempotents). The second two products given by mutual left and right multiplication of \textit{G} and \textit{H} due to Eq.(\ref{GrindEQ__3_}) vanish
\begin{equation} \label{GrindEQ__12_} 
GH\equiv G_{B}^{A} H_{C}^{B} \equiv a^{A} a_{B} b^{B} b_{C} =0,  HG\equiv H_{B}^{A} G_{C}^{B} \equiv b^{A} b_{B} a^{B} a_{C} =0,  
\end{equation} 
i.e. these matrices are orthogonal. 

\noindent Thus tensor multiplication of the basic vectors on a 2D surface yields four objects \textit{E}, \textit{F} and\textit{ G}, \textit{H}; summarize their algebraic properties. The objects \textit{E}, \textit{F} are non-symmetric, traceless $2\times 2$-matrices with zero determinants; they have zero squares (hence, norms) thus belonging to the nilpotent set, and their mutual left and right multiplication gives birth to a new couple \textit{G} and \textit{H}, objects of mixed covariance, each composed as a tensor product of same basic vector either \textit{a }or \textit{b}. The objects \textit{G}, \textit{H} are symmetric $2\times 2$-matrices with unit trace and zero determinant; their squares (hence any natural power) returns themselves thus referring the matrices to idempotent set, and they are mutually orthogonal since their left and right products vanish. One notes that the objects \textit{E }and \textit{F} possess properties characteristic to vector unit of the HP numbers subset, dual (hyperbolic) numbers$^{10}$; among the four objects they should be considered ``more fundamental'' ones since they are used to compose idempotents, not vice versa.
\section*{\small 3. SIMPLEST LINEAR COMBINATIONS OF NILPOTENTS AND IDEMPOTENTS}
\vskip-0.7em
\noindent Simplest linear combinations of the nilpotent couple \textit{E}, \textit{F} are\footnote{The tilde over symbol I  (and later over K) will be explained below.}
\begin{equation} \label{GrindEQ__13_} 
\tilde{I}_{B}^{A} \equiv E^{A} _{B} +F^{A} _{B} =a^{A} b_{B} +b^{A} a_{B} ,  
\end{equation} 
\begin{equation} \label{GrindEQ__14_} 
J^{A} _{B} \equiv E^{A} _{B} -F^{A} _{B} =a^{A} b_{B} -b^{A} a_{B} . 
\end{equation} 
As above examine algebraic properties and all possible products of the objects (\ref{GrindEQ__13_}), (\ref{GrindEQ__14_}).

\noindent The object (\ref{GrindEQ__13_}) is a symmetric traceless matrix with the determinant 
\[\det (\tilde{I}_{B}^{A} )=\frac{1}{2} (\delta _{A}^{B} \delta _{C}^{D} -\delta _{A}^{D} \delta _{C}^{B} )(a^{A} b_{B} +b^{A} a_{B} )(a^{C} b_{D} +b^{C} a_{D} )=-1.\] 
Due to Eqs.(\ref{GrindEQ__7_}), (\ref{GrindEQ__8_}) the square of $\tilde{I}$yields the sum of idempotents
\begin{equation} \label{GrindEQ__15_} 
\tilde{I}^{2} \equiv \tilde{I}_{B}^{A} \tilde{I}_{C}^{B} \equiv (E^{A} _{\;\;B} +F^{A} _{\;\;B} )(E^{B} _{\;\;C} +F^{B} _{\;\;C} )=G_{C}^{A} +H_{C}^{A} =a^{A} a_{C} +b^{A} b_{C} ,  
\end{equation} 
i.e. $\tilde{I}^{2} $ is a symmetric matrix that due to Eq.(\ref{GrindEQ__2_}) has the non-zero trace  
\begin{equation} \label{GrindEQ__16_} 
Tr(\tilde{I}^{2} )=2,   
\end{equation} 
and the determinant equal to unity
\begin{equation} \label{GrindEQ__17_} 
\det (\tilde{I}^{2} )=\frac{1}{2} (\delta _{A}^{B} \delta _{C}^{D} -\delta _{A}^{D} \delta _{C}^{B} )(a^{A} a_{B} +b^{A} b_{B} )(a^{C} a_{D} +b^{C} b_{D} )=1.  
\end{equation} 
The square of the matrix (\ref{GrindEQ__15_}) returns the initial object
\[(\tilde{I}^{2} )^{2} =(G_{B}^{A} +H_{B}^{A} )(G_{C}^{A} +H_{C}^{A} )=a^{A} a_{C} +b^{A} b_{C} =\tilde{I}^{2} , \] 
the last property and Eqs.(\ref{GrindEQ__16_}), (\ref{GrindEQ__17_}) uniquely identifying $\tilde{I}^{2} $with  the  2D Kronecker delta
\begin{equation} \label{GrindEQ__18_} 
a^{A} a_{B} +b^{A} b_{B} =\delta _{B}^{A} \equiv 1.  
\end{equation} 
Therefore Eq.(\ref{GrindEQ__15_}) written symbolically 
\begin{equation} \label{GrindEQ__19_} 
\tilde{I}^{2} =1 
\end{equation} 
means that the object $\tilde{I}_{B}^{A} $ can be thought of, apart from $\delta _{B}^{A} $, as another type of real unit in the $2\times 2$- matrix set. 
\noindent The object (\ref{GrindEQ__14_}) is a skew-symmetric traceless matrix with the determinant
\[\det (J^{A} _{B} )=\frac{1}{2} (\delta _{A}^{B} \delta _{C}^{D} -\delta _{A}^{D} \delta _{C}^{B} )(a^{A} b_{B} -b^{A} a_{B} )(a^{C} b_{D} -b^{C} a_{D} )=1.\] 
The square of the object $J$ yields the negative sum of idempotents
\[J^{2} \equiv J^{A} _{B} J^{B} _{C} =(E^{A} _{B} -F^{A} _{B} )(E^{B} _{C} -F^{B} _{C} )=-(G_{C}^{A} +H_{C}^{A} )=-(a^{A} a_{C} +b^{A} b_{C} )=-\delta _{C}^{A} , \] 
or symbolically 
\begin{equation} \label{GrindEQ__20_} 
J^{2} =-1.  
\end{equation} 
Eq.(\ref{GrindEQ__20_}) means that $J$ is, apart from $i\delta _{B}^{A} $,  a type of imaginary unit in the $2\times 2$-matrix set.

\noindent Now find that product of $J$ and $\tilde{I}$ yields difference of idempotents
\begin{equation} \label{GrindEQ__21_} 
\tilde{K}\equiv \tilde{K}_{C}^{A} \equiv J\cdot \tilde{I}=J^{A} _{B} \tilde{I}_{C}^{B} =(a^{A} b_{B} -b^{A} a_{B} )(a^{B} b_{C} +b^{B} a_{C} )=G_{C}^{A} -H_{C}^{A} =a^{A} a_{C} -b^{A} b_{C} .  
\end{equation} 
The object (\ref{GrindEQ__21_}) is a symmetric traceless matrix with the determinant
\begin{equation} \label{GrindEQ__22_} 
\det (\tilde{K}_{C}^{A} )=\frac{1}{2} (\delta _{A}^{B} \delta _{C}^{D} -\delta _{A}^{D} \delta _{C}^{B} )(a^{A} a_{B} -b^{A} b_{B} )(a^{C} a_{D} -b^{C} b_{D} )=-1;  
\end{equation} 
its square is
\[\tilde{K}^{2} \equiv \tilde{K}_{B}^{A} \tilde{K}_{C}^{B} =(G_{B}^{A} -H_{B}^{A} )(G_{C}^{B} -H_{C}^{B} )=G_{C}^{A} +H_{C}^{A} =a^{A} a_{C} +b^{A} b_{C} =\delta _{C}^{A} ,\] 
or
\begin{equation} \label{GrindEQ__23_} 
\tilde{K}^{2} =1.  
\end{equation} 
The properties (\ref{GrindEQ__22_}), (\ref{GrindEQ__23_}) of the object $\tilde{K}$ are similar to those of the object $\tilde{I}$, this means that $\tilde{K}$ represents another type of real unit. The transposition of the multipliers of Eq.(\ref{GrindEQ__21_}) gives negative expression for the same object $\tilde{K}$
\[\tilde{I}\cdot J=\tilde{I}_{B}^{A} J^{B} _{C} =(a^{A} b_{B} +b^{A} a_{B} )(a^{B} b_{C} -b^{B} a_{C} )=-(a^{A} a_{C} -b^{A} b_{C} )=-\tilde{K}_{C}^{A} \equiv -\tilde{K},\] 
the set of simplest linear combinations is complete.
\section*{\small 4. HYPERCOMPEX UNITS AND STRUCTURE OF SPACE DIMENSIONS}
\vskip-0.7em
\noindent So the four linear combinations are unit-like objects in the set of $2\times 2$-matrices. The sum of nilpotents (\ref{GrindEQ__13_}) gives a type of real unit $\tilde{I}$, the difference of nilpotents (\ref{GrindEQ__14_}) gives an imaginary unit $J$, the difference of idempotents (\ref{GrindEQ__21_}) again gives a type of real unit $\tilde{K}$; squares of these unit-like objects are expressed through the real unit 1 given by the sum of idempotents (\ref{GrindEQ__18_}). It is easily verified that products of 1 and any of the other unit-like objects return this object, e.g.
\[\tilde{K}_{B}^{A} (a^{B} a_{D} +b^{B} b_{D} )=(a^{A} a_{B} -b^{A} b_{B} )(a^{B} a_{D} +b^{B} b_{D} )=a^{A} a_{D} -b^{A} b_{D} =\tilde{K}_{D}^{A} , \] 
or symbolically $\tilde{K}\cdot 1=\tilde{K}$. But multiplication of the unit-like objects yields diversity of results dependent not only on names (hence, structure) of the units but too on their order, e.g. as in Eq.(\ref{GrindEQ__21_}) or
\[J\cdot \tilde{K}=J_{B}^{A} \tilde{K}^{B} _{C} =(a^{A} b_{B} -b^{A} a_{B} )(a^{B} a_{C} -b^{B} b_{C} )=-(a^{A} a_{C} +b^{A} b_{C} )=-\tilde{I}_{C}^{A} \equiv -\tilde{I},\] 
\[\tilde{K}\cdot J=\tilde{K}^{A} _{B} \, J_{C}^{B} =(a^{A} a_{B} -b^{A} b_{B} )(a^{B} b_{C} -b^{B} a_{C} )=a^{A} b_{C} +b^{A} a_{C} =\tilde{I}_{C}^{A} \equiv \tilde{I}.\] 
The complete multiplication table for all four 2D-vector-born units is\footnote{The units are multiplied in the order ``row by column'' with the products at the respective intersections.} 

\vskip1em
\begin{tabular}{|p{0.4in}|p{0.4in}|p{0.4in}|p{0.4in}|} \hline 
1 & $\tilde{I}$ & $J$ & $\tilde{K}$ \\ \hline 
$\tilde{I}$ & 1 & -$\tilde{K}$ & -$J$ \\ \hline 
$J$ & $\tilde{K}$ & -1 & -$\tilde{I}$ \\ \hline 
$\tilde{K}$ & $J$ & $\tilde{I}$ & 1 \\ \hline 
\end{tabular}$\;\;$. \hspace{17.5em}(24)
\vskip1em

\noindent This means that the objects $(1;\tilde{I},J,\, \tilde{K})$ behave as a set of some hypercomplex units, and the table (\ref{GrindEQ__23_}) may be regarded as a basis of a hypercomplex algebra; but even with only real coefficients at the units this algebra would comprise zero-divisors, e.g.
\[
S\equiv 1+\tilde{I}, \left|S\right|^{2} =S\cdot S^{*}\equiv (1+\tilde{I})(1-\tilde{I})=0;
\]
\vspace{-1.5em}
\[P\equiv \tilde{I}+J, P^{2} \equiv P\cdot P=(\tilde{I}+J)(\tilde{I}+J)=0,
\] 
the number \textit{S}  belonging to the set of split-complex (double) numbers, the number \textit{P} being a dual number$^{10}$. 

\noindent However one easily changes the ``non-satisfactory'' set of units $(1;\, \, \tilde{I},J,\, \tilde{K})$ so that it becomes the basis of ``good'' quaternion algebra. In fact only the objects with the tilde$\, \tilde{I}$,$\, \tilde{K}$ are to be converted from real-like units into imaginary ones (in the set of $2\times 2$-matrices) what is done with the help of the scalar imaginary unit \textit{i}. Then the linear combinations of products of orthonormal vectors belonging to a 2D-surface give a set of quaternion units 
\[
1\equiv \delta _{B}^{A} =a^{A} a_{B} +b^{A} b_{B},
\tag{25a}
\]
\[
I\equiv I_{B}^{A} =-i(a^{A} b_{B} +b^{A} a_{B} ),
\tag{25b}
\]
\[
J\equiv J^{A} _{\;\;B} =a^{A} b_{B} -b^{A} a_{B},
\tag{25c}
\]
\[
K\equiv K_{B}^{A} =i\, (a^{A} a_{B} -b^{A} b_{B} ),
\tag{25d}
\]
\noindent original Hamilton's notations (with no tildes) used. The units (25) form the standard quaternion multiplication table
\vskip1em
\begin{tabular}{|p{0.3in}|p{0.4in}|p{0.4in}|p{0.4in}|} \hline 
1 & \textit{I} & \textit{J} & \textit{K} \\ \hline 
\textit{I} & -1 & \textit{K} & \textit{-J} \\ \hline 
\textit{J} & \textit{-K} & -1 & \textit{I} \\ \hline 
\textit{K} & \textit{J} & \textit{-I} & -1 \\ \hline 
\end{tabular}$\;\;.$ \hspace{18em}(26)
\vskip1em

\noindent Transition to 3D vector notation $(I,J,\, K)$ $\rightarrow$ $(\, \textbf{q}_{1} ,\, \textbf{q}_{2} ,\textbf{q}_{3} )=\textbf{q}_{k} $, $k,n,l...=1,\, 2,\, 3$ shrinks the table (26) to the familiar compact form 
\setcounter{equation}{26}
\begin{equation} \label{GrindEQ__27_} 
1\cdot \textbf{q}_{k} =\, \textbf{q}_{k} \cdot 1\equiv \, \textbf{q}_{k} , \quad \textbf{q}_{k} \textbf{q}_{n} =-\delta _{kn} \, +\varepsilon _{knj} \textbf{q}_{j} ,  
\end{equation} 
$\delta _{kn} \, ,\, \, \varepsilon _{knj} $ being 3D Kronecker and Levi-Civita symbols, summation rule still valid. Geometrically the units $\textbf{q}_{k} $ behave as three vectors initiating a Cartesian Q-frame in 3D space, each unit determining one dimension. So Eqs.(25) reveal the non-obvious fact that each dimension of the 3D space (physical space not excluded) may be thought of having some ``fine structure'', the structural elements reflecting geometric properties of some two-dimensional space (surface). This means that there is a clear functional interdependence between geometry of a surface and behavior of respective frame; therefore each particular frame (with an observer implied in its origin) immanently has its ``elementary image'' represented by a couple of vectors forming a 2D surface domain, a ``2D-sell''.
\section*{\small 5. THE METRICS OF 2D-SELLS ASSOCIATED WITH THE Q-UNITS}
\vskip-0.7em
\noindent The above study offers solution of ``direct problem'', of building a Q-frame from basic 2D vectors. Procedure of the ``inverse problem'' solution, of describing the 2D-sell on the base of properties of a given Q-frame, is prompted by Eqs.(25). Select e.g. Eq.(25c)  with the identification $\textbf{q}_{3} \equiv K_{B}^{A} $, and notice that this matrix has two eigenvectors $a^{B} ,b^{B} $ and respective covectors $a_{A} ,b_{A} $ with eigenvalues $\pm i\, $
\begin{equation} \label{GrindEQ__28_} 
K_{B}^{A} a^{B} =i\, a^{B} , K_{B}^{A} a_{A} =i\, a_{B} ; \quad K_{B}^{A} b^{B} =-i\, b^{B} , K_{B}^{A} b_{A} =-i\, b_{B} .   
\end{equation} 
If the eigenfunctions are known, then according to Eqs.(24, 25) all other Q-units, the scalar one included, are straightforwardly constructed. This moves out the eigenfunction problem for given Q-vector triad, and makes it interesting to express eigenvectors belonging to other two Q-units (25a) and (25b) through vector and covector solutions of Eqs.(\ref{GrindEQ__28_}). The procedure of finding the expressions is straightforward. Each eigenvector is assumed to be a linear combinations of the vectors $a^{B} $, $b^{B} $, the coefficients of the combination determined from respective eigenfunction equation; in fact the solutions for the functions are found up to an arbitrary factor. A solution for eigenfunctions of the operator $\textbf{q}_{1} \equiv I_{B}^{A} $ is (with the free factor chosen a constant) 

\[
\mbox{for $+i\, $}: \displaystyle c^{A} \equiv \frac{i}{\sqrt{2} } (a^{A} -b^{A} ),\, c_{A} \equiv -\frac{i}{\sqrt{2} } (a_{A} -b_{A} ),
\tag{29a}
\]
\[\mbox{for $-i\, $}: \displaystyle d^{A} \equiv \frac{1}{\sqrt{2} } (a^{A} +b^{A} ),\, d_{A} \equiv \frac{1}{\sqrt{2} } (a_{A} +b_{A} ); \tag{29b}
\]
\noindent a solution for the operator $\textbf{q}_{2} \equiv J_{B}^{A} $ is
\[
\mbox{for $+i\, $}: \displaystyle e^{A} \equiv \frac{1}{\sqrt{2} } (a^{A} +ib^{A} ),\, e_{A} \equiv \frac{1}{\sqrt{2} } (a_{A} -ib_{A} ), \tag{29c}
\]
\[
\mbox{for $-i\, $}: \displaystyle f^{A} \equiv -\frac{1}{\sqrt{2} } (a^{A} -ib^{A} ),\, f_{A} \equiv -\frac{1}{\sqrt{2} } (a_{A} +ib_{A} ). \tag{29d}
\]
\setcounter{equation}{29}
\noindent Standard compositions of the type (25) of these eigenfunctions give the same set (26) of the Q-units but ``from the viewpoint'' of the unit \textit{I} or the unit \textit{J}, e.g.
\[I\equiv I_{C}^{A} =i\, (c^{A} c_{B} -d^{A} d_{B} ), J\equiv J_{C}^{A} =i\, (e^{A} e_{B} -f^{A} f_{B} ). \] 
Analysis of the quaternion multiplication table (26), (\ref{GrindEQ__27_}) shows$^{12}$ that the vector Q-units' eigenfunctions $a^{B} $, $b^{B} $ and those given by Eqs.(29) has spinor properties since the multiplication table (26) [or (\ref{GrindEQ__27_})] remains invariant under transformations of the eigenfunctions by $2\times 2$-matrices from special linear group $SL(2,C)$. 

\noindent According to results of the Section 4 each Q-unit should have an associated 2D-sell with a metric constructed from the Q-unit's eigenfunctions. Find expression for all these metrics in the terms of the eigenfunctions and also in terms of the initially chosen spinors $a_{A} $, $b_{A} $. First, find expression for metric $g_{(\ref{GrindEQ__3_})AB} $ of the 2D-sell associated with $\textbf{q}_{3} $ through its eigenfunctions $a_{A} $, $b_{A} $; the sought for expression immediately follows from Eqs. (\ref{GrindEQ__18_}) or (25a)
\vskip-1.5em
\begin{equation} \label{GrindEQ__30_} 
g_{(3)AB} =a_{A} a_{B} +b_{A} b_{B} ,  
\end{equation} 
the orthonormality conditions (\ref{GrindEQ__2_}) and (\ref{GrindEQ__3_}) evidently fulfilled. The formula for metric $g_{(\ref{GrindEQ__1_})AB} $ associated with $\textbf{q}_{1} $ is given by analogous combination of the covector-eigenfunctions $c_{A} ,\, \, d_{B} $ and with the help of Eqs. (29a), (29b) is expressed through $a_{A} $, $b_{A} $
\vskip-1.5em
\begin{equation} \label{GrindEQ__31_} 
g_{(1)AB} =c_{A} c_{B} +d_{A} d_{B} =a_{A} b_{B} +a_{B} b_{A} .  
\end{equation} 
One immediately notices that Eq.(\ref{GrindEQ__31_}) is equivalent to Eq.(\ref{GrindEQ__13_}) with its upper index lowered by metric (\ref{GrindEQ__30_}), i.e. the metric of the 2D-sell belonging to unit Q-vector $\textbf{q}_{1} $ in terms of basic elements $a_{A} $, $b_{A} $ belonging to $\textbf{q}_{3} $ is perceived as the vector real unit $\tilde{I}_{AB} $ with all lower indices. 

\noindent The metric associated with $\textbf{q}_{2} $ is found through its eigenfunctions, and with the help of Eqs.(29c), (29d) through those belonging to $\textbf{q}_{3} $, as
\vskip-1.5em
\begin{equation} \label{GrindEQ__32_} 
g_{(2)AB} =e_{A} e_{B} +f_{A} f_{B} =a_{A} a_{B} -b_{B} b_{A} ,  
\end{equation} 
i.e. in terms of $a_{A} $, $b_{A} $ this metric is perceived as vector real unit $\tilde{K}_{AB} $ with all lower indices. 

\noindent Eqs.(\ref{GrindEQ__31_}), (\ref{GrindEQ__32_}) to some extent reveal the geometric sense of the real vector units $\tilde{I}$, $\tilde{K}$ emerging in Eqs.(\ref{GrindEQ__13_}), (\ref{GrindEQ__21_}) as simple linear combinations of nilpotents and idempotents. If Kronecker delta (\ref{GrindEQ__18_}) is a ``natural'' metric of a 2D-sell (the metric of the surface structuring a certain, say basic, space dimension) then the units $\tilde{I}$and$\tilde{K}$ are similar metrics of two other space dimensions but regarded from viewpoint of the basic dimensions.

\section*{\small 6. ``WORLD SCREEN TECHNOLOGY'' AND EXAMPLES OF Q-FRAMES BORN BY GIVEN 2D-SELLS}
\vskip-0.7em
\noindent A quaternion triad is known to be naturally associated with a frame of reference$^8$ providing exhaustive explanation of motion of an arbitrary particle. But as is shown above such a frame has its ``more elementary'' image, a 2D-sell formed by a couple of 2D eigenvectors belonging to any of three unit Q-vectors. Therefore characteristics of the particle motion are reflected by mathematical properties of the 2D-sell description. Let motion law of a certain number of particles be known; then the chosen space domain with the particles in it can be adequately represented by a regular set of respective 2D-sells each sell describing behavior of one particle. A collection of all such sells form a kind of screen containing full kinematical information of the domain of 3D world. Eqs.(29-32) state that there are at least three options to construct the ``world screen'' using different sets of eigenfunctions (however all linearly dependent). Detailed analysis of the 2D-sells reflecting properties of characteristic 3D motions is to appear in following publications.

\noindent Now stress that, vice versa, an arbitrary 2D surface (its sufficiently limited domain) as well should generate a Q-frame. Some illustrative examples of Q-frames born by ``ordinary'' geometric surfaces are given below.

\noindent \underbar{Plane}. This trivial example details the procedure of constructing Q-frames from 2D-sells. The Cartesian metric of a plane assumed structuring the dimension $\textbf{q}_{3} $ is the Kronecker delta $g_{(\ref{GrindEQ__3_})AB} =\delta _{A1} \delta _{B1} +\delta _{A2} \delta _{B2} $, $g_{(\ref{GrindEQ__3_})} ^{AB} =\delta _{1}^{A} \delta _{1}^{B} +\delta _{2}^{A} \delta _{2}^{B} $,  
\noindent in this case the basic vectors forming the 2D-sell are constant rows and columns (plane spinors)
\[a_{A} =\delta _{A2} =\left(\begin{array}{cc} {0} & {1} \end{array}\right),\, \, \, b_{A} =\delta _{A1} =\left(\begin{array}{cc} {1} & {0} \end{array}\right);  a^{A} =\delta _{2}^{A} =\left(\begin{array}{c} {0} \\ {1} \end{array}\right),\, \, b^{A} =\delta _{1}^{A} =\left(\begin{array}{c} {1} \\ {0} \end{array}\right).\] 
Using Eqs.(25) build the vector Q-units

\begin{eqnarray}
\textbf{q}_{\tilde{3}}&=&i\, (a^{A} a_{B} -b^{A} b_{B} )=-i\, \left(\begin{array}{cc} {1} & {0} \\ {0} & {-1} \end{array}\right),\nonumber\\
\textbf{q}_{\tilde{1}}&=&-i(a^{A} b_{B} +b^{A} a_{B} )=-i\, \left(\begin{array}{cc} {0} & {1} \\ {1} & {0} \end{array}\right),\nonumber\\
\textbf{q}_{\tilde{2}}&=&a^{A} b_{B} -b^{A} a_{B}=-i\, \left(\begin{array}{cc} {0} & {-i} \\ {i} & {0} \end{array}\right),
\end{eqnarray}
\noindent having here the canonical form (the Pauli matrices with \textit{--i} factor). The set (33) of constant Q-units describes an inertial frame.

\noindent \underbar{Cylinder}. The metric (again associated with $\textbf{q}_{3} $) has the form 
\[g_{(3)AB} =\delta _{A1} \delta _{B1} +e^{2\eta } \delta _{A2} \delta _{B2} , g_{(3)} ^{AB} =\delta _{1}^{A} \delta _{1}^{B} +e^{-2\eta } \delta _{2}^{A} \delta _{2}^{B} ,\] 
non-zero (and unit-free) radius of the cylinder for convenience is written as $r\equiv e^{\eta } $, $\eta $ being a real constant. The basic covectors and vectors of the cylinder 
\[a_{A} =e^{\eta } \delta _{A2} ,\, \, \, b_{A} =\delta _{A1} ;  a^{A} =e^{-\eta } \delta _{2}^{A} ,\, \, b^{A} =\delta _{1}^{A} \] 
when substituted to Eqs (25) yield the vector Q-units

\begin{equation} \label{GrindEQ__34_} 
\textbf{q}_{3} =-i\, \left(\begin{array}{cc} {1} & {0} \\ {0} & {-1} \end{array}\right), \textbf{q}_{1} =-i\, \left(\begin{array}{cc} {0} & {e^{\eta } } \\ {e^{-\eta } } & {0} \end{array}\right), \textbf{q}_{2} =-i\, \left(\begin{array}{cc} {0} & {-ie^{\eta } } \\ {ie^{-\eta } } & {0} \end{array}\right).  
\end{equation} 
The units (\ref{GrindEQ__34_}) are well known to emerge as a result of the simple hyperbolic rotation $\textbf{q}_{k} =H_{k\tilde{n}} \textbf{q}_{\tilde{n}} $ with
\begin{equation} \label{GrindEQ__35_} 
H_{k\tilde{n}} =\, \left(\begin{array}{ccc} {\cosh \eta } & {-i\, \sinh \eta } & {0} \\ {i\sinh \eta } & {\cosh \eta } & {0} \\ {0} & {0} & {1} \end{array}\right),  
\end{equation} 
$\textbf{q}_{\tilde{k}} $ given by Eqs.(33). Quaternion version of relativity theory$^8$ states that the units (\ref{GrindEQ__34_}) describe a frame moving with constant velocity\footnote{Fundamental velocity is chosen a unity.} $V=\tanh \eta $ in the positive sense of the direction $\textbf{q}_{2} $ and being observed from the immobile frame $\textbf{q}_{\tilde{k}} $; the time coordinate in this case changes along imaginary direction  $i\, \textbf{q}_{1} =\tilde{I}$. It is evident that a cylindrical surface with changing radius corresponds to a relativistic frame having rectilinear trajectory but variable velocity modulus. 

\noindent \underbar{Sphere}. The spherical metric (its reciprocal) is a clear composition of the covectors (vectors)
\[a_{A} =r\, \sin \vartheta \, \delta _{A2} ,\, \, \, b_{A} =r\, \delta _{A1} ;  \quad a^{A} =\frac{1}{r\, \sin \theta } \delta _{2}^{A} ,\, \, b^{A} =\frac{1}{r} \delta _{1}^{A} \] 
that due to Eqs.(25) give birth to the vector Q-units
\begin{equation} \label{GrindEQ__36_} 
\textbf{q}_{3} =-i\, \left(\begin{array}{cc} {1} & {0} \\ {0} & {-1} \end{array}\right), \textbf{q}_{1} =-i\, \left(\begin{array}{cc} {0} & {\sin \theta } \\ {\sin ^{-1} \theta } & {0} \end{array}\right), \textbf{q}_{2} =-i\, \left(\begin{array}{cc} {0} & {-i\sin \theta } \\ {i\sin ^{-1} \theta } & {0} \end{array}\right) 
\end{equation} 
having in this case a singularity at the polar point. To avoid infinities consider a domain close to the sphere's equator $\displaystyle \theta =\frac{\pi }{2} -\sigma $, $\sigma \ll 1$, so that $\displaystyle \sin \theta =\cos \sigma \cong 1-\frac{\sigma ^{2} }{2} $; then one obtains a development of e.g. vector $i\textbf{q}_{1} $ of Eqs.(\ref{GrindEQ__36_}) 
\[\textbf{q}_{1} \cong \textbf{q}_{\tilde{1}} -i\, \frac{\sigma ^{2} }{2} \textbf{q}_{\tilde{2}} \] 
describing hyperbolic rotation of the frame (\ref{GrindEQ__36_}) similar to that given by Eq.(\ref{GrindEQ__35_}) but with small hyperbolic parameter $\displaystyle \eta =\frac{\sigma ^{2} }{2} \ll 1$. Thus the spherical ring corresponds to a frame as in the previous case moving along $\textbf{q}_{2} $ relatively to $\textbf{q}_{\tilde{k}} $ but with variable velocity $\displaystyle V=\tanh \eta \cong \frac{\sigma ^{2} }{2} $. Other characteristic examples will be given elsewhere.

\section*{\small 7. DISCUSSION}
\vskip-0.7em
\noindent A somewhat prolonged and dull study undertaken in the first sections of this paper of nilpotent and idempotent objects emerging upon 2D surface geometry has nonetheless an exciting issue. A set of strict mathematical correlations is obtained linking geometry of 2D surface domains with functional dependence of 3D hypercomplex units. Existence of these math links, on the one hand, may correspond to presence of real space (and time) dimensions' interior structure not detected in macroscopic experiments. And since the basic vectors forming elementary 2D surfaces behave like spinors this aspect of the study may lead to better understanding of essence of the quantum theory. On the other hand, the found correlations definitely open opportunity to replace description of quaternion triads successfully treated in physics as movable rigid frames of reference by description of respective 2D-sells geometry. Moreover it turns out that any Q-frame, having a sense of an ``oriented particle'' (body of reference), can be adequately represented by at least three different 2D-sells each associated with a chosen space dimension. This prompts to suggest an idea of mapping domains of 3D space comprising a number of particles (Q-frames) onto a sufficiently wide ``world screen'' consisting of respective number of 2D-sells, each sell represented by a somehow limited local ``spinor-screen''. Few given examples show that the idea of the ``screen'' is in principle realizable although a good deal of profound study remains yet undone, so there is a space for explorations aiming to make this alluring technology work and as well be useful. 

\pagebreak

\noindent \textbf{References and Notes}

\begin{enumerate}
\item J. P. Ward, Quaternions and Cayley Numbers: Algebra and Applications, Kluwer Academic Publ., Dordrecht and Boston \textbf{(1997)}

\item V.Trifonov, Int.J.Theor.Phys.V46, No2, 251 \textbf{(2007)}

\item Ch.F.F.Karney,  J.Mol.Graph.Mod. V25, No 5, 595 \textbf{(2007)}

\item  A.Gsponer, J.-P.Hurni, Quaternions in mathematical physics (\ref{GrindEQ__2_}): Analytical bibliography, arXiv:math-ph/0510059v4 6 July \textbf{2008 }

\item L.P.Horovitz, L.C.Beidenharn, Annales of Physics, V157, 432 \textbf{(1984)}

\item S.L.Adler, Quaternionic quantum mechanics and quantum fields, Oxford Univ.Press., N.Y. \textbf{(1995)}

\item  N.V.Mitskievich, Physicfl Fields in the Theory of Relativity, Moscow, Nauka publ. (\textbf{1969)} 

\item  A.P.Yefremov, Adv.Sci.Lett., V1,179 \textbf{(2008) }

\item A.P.Yefremov, Quaternionic Program. Generalized Theories and Experiments, Kluwer Acad. Publ., Netherlands, p. 395-409 \textbf{(2004) }

\item A.P.Yefremov, Adv.Sci.Lett., V3,537 \textbf{(2010) }

\item P.Rastall, Rev. Mod. Phys., V2, 820 \textbf{(1964) }

\item A.P.Yefremov, Quaternion and Biquaternions: Algebra, Geometry and Physical\textbf{ }Theories, arXiv:math-ph/0501055v1, January \textbf{(2005) }

\item A.P.Yefremov, Gravitation and Cosmology, V16, No 2, 137 \textbf{(2010) }
\end{enumerate}

\end{document}